\documentclass[final,twocolumn,authoryear]{elsarticle}

\usepackage{graphicx}

\usepackage{color}
\usepackage{graphicx}
\usepackage{amssymb}

\journal{Planetary and Space Science}

\begin{document}

\begin{frontmatter}



\title{Scaling of convective velocity in a vertically vibrated granular bed}


\author{Tomoya M. Yamada\footnote{Corresponding author. Tel.:+41 52 789 3014; fax: +81 52 789 3013. E-mail address: yamada.tomoya@a.mbox.nagoya-u.ac.jp} and Hiroaki Katsuragi}

\address{Department of Earth and Environmental Sciences, Nagoya University, Furocho, Chikusa, Nagoya 464-8601, Japan}

\begin{abstract}
We experimentally study the velocity scaling of granular convection which is a possible mechanism of the regolith migration on the surface of small asteroids. In order to evaluate the contribution of granular convection to the regolith migration, the velocity of granular convection under the microgravity condition has to be revealed. Although it is hard to control the gravitational acceleration in laboratory experiments, scaling relations involving the gravitational effect can be evaluated by systematic experiments. Therefore, we perform such a systematic experiment of the vibration-induced granular convection. From the experimental data, a scaling form for the granular convective velocity is obtained. The obtained scaling form implies that the granular convective velocity can be decomposed into two characteristic velocity components: vibrational and gravitational velocities. In addition, the system size dependence is also scaled. According to the scaling form, the granular convective velocity $v$ depends on the gravitational acceleration $g$ as $v \propto g^{0.97}$ when the normalized vibrational acceleration is fixed.
\end{abstract}

\begin{keyword}
regolith migration, granular convection, scaling analysis, gravitational acceleration.

\end{keyword}

\end{frontmatter}


\section{Introduction}
\label{sec:introduction}
In the solar system, there are many small bodies such as asteroids and comets. Since these small astronomical objects could keep the ancient information of the history of the solar system, a lot of efforts have been devoted to investigations of these objects so far~(e.g., \citet{AsteroidsIII}). For instance, the asteroid Itokawa was explored by the Japanese space craft {\it Hayabusa} from September 2005. The exploration uncovered the details of Itokawa's surface terrain. Itokawa is covered by various sizes of granular matter such as regolith, pebbles, and boulders~\citep{Fujiwara2006}. Moreover, migrations and sorting of the regolith could occur on the surface of Itokawa~\citep{Miyamoto2007}, although the surface gravity of Itokawa is extremely low. Impact craters located on the surface of Itokawa were very subtle, i.e., they show indistinct morphologies~\citep{Saito2006,Hirata2010}. This is probably due to the erasure of the craters by seismic shaking~\citep{Michel2009}. In addition, tiny samples returned from Itokawa allow us to analyze the detail of its history. \citet{Nagao2011} revealed that Itokawa's surface grains are relatively young in terms of cosmic-rays exposure. The estimated age is approximately eight million years. Besides, using x-ray microtomography,~\citet{Tsuchiyama2011} found that some particles had rounded edges. All these facts suggest that the surface of Itokawa would be active and continue to be renewed until recently. One possible explanation of such young surface is the regolith convection caused by impact-induced seismic shaking.

Although the direct measurement of the seismic wave has not been achieved, \citet{Richardson2004} and \citet{Richardson2005} studied the possibility of global seismic shaking of asteroid Eros in order to explain its surface modification processes. They built a model of seismic shaking by considering the attenuating diffusion of the seismic wave. \citet{Miyamoto2007} partially applied the model to the asteroid Itokawa and showed that the global regolith convection might occur even by small-scale impacts. To unlock the regolith grains network supported by gravity, the maximum acceleration induced by the seismic shaking has to be greater than the gravitational acceleration. \citet{Miyamoto2007} revealed that this criterion can be satisfied by a small-scale impact since the gravity on the surface of Itokawa $g_I$ is very small, $g_I \simeq 10^{-4}$ m/s$^2$~\citep{Abe2006,Fujiwara2006}. The value of $g_I$ is about five orders of magnitude smaller than the Earth's gravitational acceleration, $g_E=9.8$ m/s$^2$. The evaluation is still qualitative since they only assessed the onset criterion of the regolith convection. While the quantitative assessment of the convective velocity is necessary to discuss the feasibility of the regolith convection, there have been very few such studies.

Granular convection can be generally observed in a granular matter under the mechanical vibration. When a granular matter is subjected to a steady vertical vibration, granular convection is readily induced. If the vibrated granular matter is polydisperse, the size segregation of grains occurs usually in vertical direction. This vibration-induced size segregation is called Brazil nut effect (BNE). The BNE can be caused by the granular convection~\citep{Knight1993}. This means that the grains can be migrated and sorted simultaneously by vibration. Although the regolith convection accompanied by migration and sorting seems to be a natural outcome of global seismic shaking, the scaling approach to the granular convective velocity is needed to discuss the consistency between regolith convection and observational data such as surface age of the asteroid. Therefore, we experimentally measure and scale the velocity of granular convection in this study, as a first-step approach to this problem.

Fundamental nature of granular convection itself is an intriguing problem. The granular convection has long been studied both by experiments~\citep{Faraday1831,Ehrichs1995,Knight1996,Pastor2007,Garcimartin2002} and numerical simulations~\citep{Taguchi1992, Luding1994, Tancredi2012}. The onset of granular convection depends on the maximum acceleration of the applied vibration. If the maximum acceleration is less than the gravitational acceleration, any granular convection does not occur at all~(e.g., \citet{Garcimartin2002}). Thus the dimensionless parameter $\Gamma$, which represents the balance between the maximum vibrational acceleration and the gravitational acceleration $g$, has been used to characterize the behavior of granular convection. $\Gamma$ is defined as, 
\begin{equation}
\Gamma = \frac{A_0 \left ( 2 \pi f \right )^{2}}{g},
\label{eq:G}
\end{equation}
where $A_0$ is the vibration amplitude and $f$ is the frequency. 

In this study, the scaling method is applied to the analysis of granular convective velocity. We are interested in asteroidal-scale granular convection while the actual experiment is limited within the laboratory scale. In such a situation, the scaling is the only way to derive a meaningful quantitative relation. In the scaling analysis, dimensionless parameters such as $\Gamma$ are useful since they do not depend on the system of unit. The weak point of the scaling analysis is its arbitrariness. Of course, $\Gamma$ is one of the most important dimensionless parameters to discuss the vibrated granular matter. However, the choice of the important dimensionless parameter is not unique. 

Specifically, another dimensionless parameter called shaking parameter $S$ was firstly introduced by~\citet{Pak1993} and recently used to describe the transitions among the granular Leidenfrost, bouncing bed, undulations, convection, and so on~\citep{Eshuis2005,Eshuis2007}. Particularly, $S$ is relevant to characterize a strongly shaken shallow granular convection~\citep{Eshuis2010}. $S$ is defined as, 
\begin{equation}
S = \Gamma \cdot \frac{A_0}{d} = \frac{\left( 2 \pi A_0 f \right)^{2} }{gd},
\label{eq:S}
\end{equation}
where $d$ is the constitutive grains diameter. $S$ denotes the balance between the squared vibrational velocity and the squared gravitational velocity. Furthermore, $S$ can be also obtained by the natural non-dimensionalization of the granular-hydrodynamic model for the strongly shaken granular convection~\citep{Eshuis2010}. Using the aforementioned dimensionless parameters, we would like to find a useful scaling relation between granular convective velocity, the gravitational acceleration, and other parameters. Therefore, we carry out a systematic laboratory experiments of the vibration-induced granular convection.

This paper is constructed by following sections. In section~\ref{sec:experiment}, we explain the experimental setup and how to measure the convective velocity. Section~\ref{sec:Results} shows the characterization of some convective-roll patterns and the result of the scaling. In section~\ref{sec:Discussion}, we discuss physical meaning and tentative implication of the scaling to the microgravity environment. Section~\ref{sec:Conclusions} contains a conclusions.

\section{Materials and Methods}
\label{sec:experiment}
\begin{figure}
\begin{center}
\scalebox{0.9}[0.9]{\includegraphics{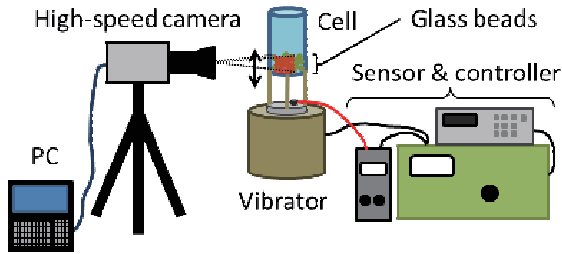}}
\end{center}
\caption{A schematic diagram of the experimental setup. Glass beads are poured into a cylindrical cell and the cell is shaken by a vibrator. Glass beads motion on the side wall is captured by a high-speed camera and the convective velocity is measured by PIV method.}
\label{fig:setup}
\end{figure}

A schematic illustration of the experimental apparatus is shown in Fig.~\ref{fig:setup}. The experimental setup consists of a cylinder made by plexiglass of its height $150$ mm and inner radius $R$ ($R=16.5$, $37.5$, or $75$ mm). Glass beads are poured in the cylindrical cell to make a granular bed of the height $H=20$, $50$, $80$, or $110$ mm. The system is mounted on an electromechanical vibrator (EMIC 513-B/A) and shaken vertically. The vibration frequency $f$ is varied from $10$ to $300$ Hz and $\Gamma$ is varied from $2$ to $6$. The grains used in this study are glass beads. Most of experiments are carried out with glass beads of diameter $d=0.8$ mm (AS-ONE corp. BZ08) and some of them are performed with glass beads of $d=0.4$ or $2$ mm (AS-ONE corp. BZ04, BZ2). Size dispersion of glass beads is less than $25$\%, and the true density of glass beads is $2.5 \times 10^3$ kg/m$^3$.  

\begin{figure*}
\begin{center}
{\includegraphics{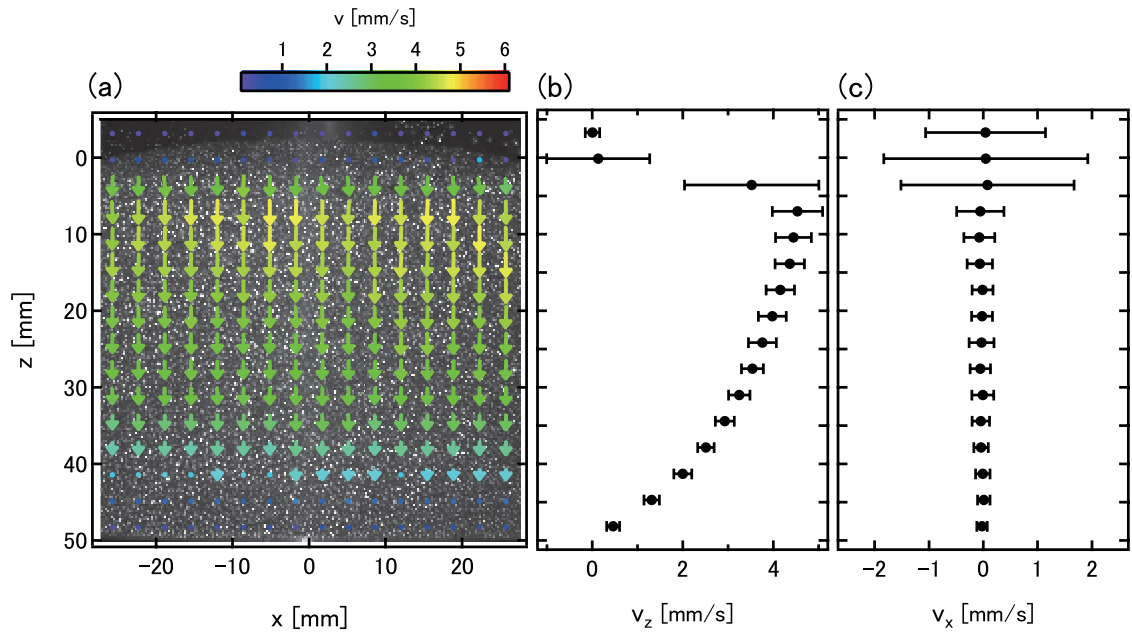}}
\caption{Convective velocity structure in the toroidal-roll (axisymmetric convection) state. Experimental conditions are $\Gamma=4$, $f=200$ Hz, $d=0.8$ mm, $H=50$ mm, and $R=37.5$ mm. (a) Each arrow length indicates the magnitude of temporarily averaged ($5.5$ s) convective velocity at each box ($64 \times 64$ pixels), which is emphasized by coloring. Colored dots indicate the small convective velocity cases (less than $2$ mm/s). The center of container corresponds to $x=0$. The convective velocity is decomposed into $z$-component $v_z$, and $x$-component $v_x$. Velocity profiles computed by horizontal average are shown in (b) $v_z$ and (c) $v_x$. Error bars in (b) and (c) are spatio-temporal standard deviations of $v_z$ and $v_x$, respectively. $v_z(z)$ shows a simple $z$-dependence (basically a decreasing function). $v_x(z)$ slightly fluctuates around $v_x(z)=0$ almost independently of $z$.}
\label{fig:vector_and_vertical_plot}
\end{center}
\end{figure*}

\begin{figure*}
\begin{center}
{\includegraphics{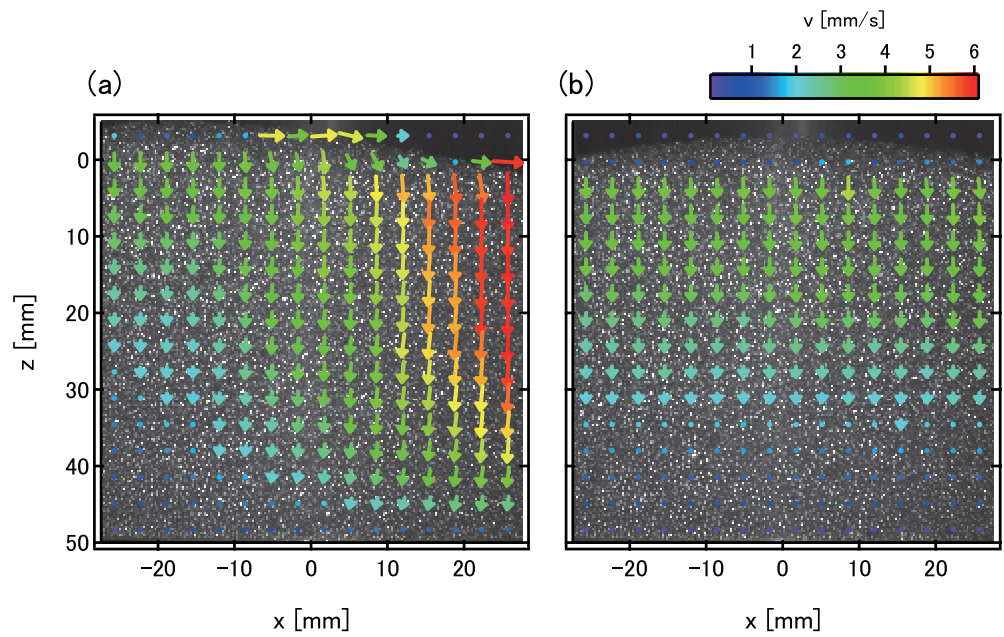}}
\caption{Convective velocity field on the side wall at (a) $\Gamma = 2$ and $f=50$ Hz, and at (b) $\Gamma = 2$ and $f=100$ Hz. Both data are taken with the granular bed of $d=0.8$ mm, $H=50$ mm, and $R=37.5$ mm. The format is same as that of Fig.~\ref{fig:vector_and_vertical_plot}(a).  In panel (a), a globally inhomogeneous velocity field can be confirmed. In contrast, rather homogeneous downward flow is observed in panel (b). The former corresponds to single-roll state and the latter corresponds to toroidal-roll state (see text and Fig.~\ref{fig:phase_diagram} for the details). 
}
\label{fig:vectorplot}
\end{center}
\end{figure*}

To measure the granular convective velocity, particle imaging velocimetry (PIV) method is utilized~\citep{Lueptow2000, Bokkers2004,Zeilstra2008}. Motions of glass beads are filmed by a high-speed camera (Photoron SA-5) through a transparent side wall. To erase the memory effect of granular matter, one minute pre-vibration is applied before each experimental realization. This means that we measure the steady granular convection. While the actual impact-induced regolith convection might be transient, we have to concentrate on the steady convection to eliminate the memory effect. The frame rate is fixed at $1,000$ fps and spatial resolution of the image ranges from $54$ $\mu$m/pixel to $130$ $\mu$m/pixel depending on the experimental conditions. The high-speed images are acquired for $5.5$ s. Since each image consists of $1,024 \times 1,024$ pixels, the size of field of view ranges from $55 \times 55$ to $133 \times 133$ mm$^2$. Raw data images are shown in backgrounds of Figs.~\ref{fig:vector_and_vertical_plot} and \ref{fig:vectorplot}. Since it is hard to completely follow all the grains' motion, we use PIV method instead of the direct particle tracking. In the analyses of Figs.~\ref{fig:vector_and_vertical_plot} and \ref{fig:vectorplot} (a), each instantaneous image is divided into $16 \times 16$ boxes, i.e., each box consists of $64 \times 64$ pixels. Note that the number of image partitioning depends on the experimental condition. Then the velocity at each box is computed by detecting a peak of the cross-correlation between two different time snapshots. Obtained (time-averaged) examples of the velocity field are shown by colored vectors in Figs.~\ref{fig:vector_and_vertical_plot} and \ref{fig:vectorplot}. An interval time to compute the velocity (cross-correlation) is kept constant so that it corresponds to a multiple of the period of vibration, i.e., two cross-correlated images are kept in same phase. For instance, $0.01$ s time interval is used for $100$ Hz vibration. If we use a full temporal resolution to calculate the velocity field, vibrational motion of individual grains can be captured just like \citet{Pastor2007} measurement. However, we are mainly interested in the global convective motion rather than such microscopic vibration. Therefore, we use phase-matching images to compute the velocity. 

As shown in Fig.~\ref{fig:vector_and_vertical_plot}, convective-velocity vectors are decomposed into $z$ (vertical) and $x$ (horizontal) directions. They are averaged along the horizontal axis since we are mainly focusing on the axisymmetric convective flow (like Fig.~\ref{fig:vector_and_vertical_plot}) for the reason mentioned later. Then we obtain vertical and horizontal components of velocities as functions of $z$, respectively as, $v_z(z)$ and $v_x(z)$. In Fig.~\ref{fig:vector_and_vertical_plot}, panels (b) and (c) respectively show $v_z(z)$ and $v_x(z)$ computed from the velocity field shown in panel (a). The positive direction of $z$ axis is taken to downward, and $z=0$ corresponds to the top surface of the granular bed. $x=0$ is the center of container. Above the surface of granular bed ($z\leqslant0$), saltations of grains are dominant and the measured velocity becomes random. This is the reason of large errors in $v_z$ and $v_x$ at $z\leqslant0$ (Fig.~\ref{fig:vector_and_vertical_plot}(b,c)). 

\section{Results and analyses}
\label{sec:Results}
\subsection{Global structure of granular convection}
\label{subsec:convective_structure}
\begin{figure}
\begin{center}
{\includegraphics{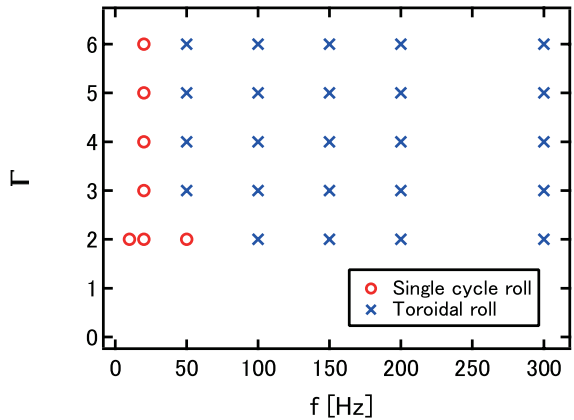}}
\caption{A phase diagram of the convective structures in $f$-$\Gamma$ space with $R=37.5$ mm, $H=50$ mm and $d=0.8$ mm. The symbols represent the observed structures of granular convection; open circles correspond to the single-roll state and cross markers indicate the toroidal-roll state. A phase boundary can be found around $f=50$ Hz independently of $\Gamma$.}
\label{fig:phase_diagram}
\end{center}
\end{figure}

First, we are going to focus on the global structure of the granular convective motion. We find that the global structure shows a transition from a single roll to a toroidal roll as $f$ increases. In the former state, grains rise up on the one side wall and fall down on the opposite side wall. Thus, in this state, the resultant velocity field shows a certain circularity. Due to this circularity, a gradient of the velocity field is observed as shown in Fig.~\ref{fig:vectorplot}(a). In this single-roll state, the degree of velocity gradient or circularity strongly depends on the viewing spot. In the latter state, on the other hand, grains rise up at the center of container and fall down on all over the wall homogeneously. The inner structure of this granular convective mode has been observed by using the magnetic resonance imaging (MRI) method~\citep{Ehrichs1995,Knight1996}. In other words, this toroidal convection state is axisymmetric. Typical examples of raw images for such homogeneous convection are shown in Figs.~\ref{fig:vector_and_vertical_plot}(a) and \ref{fig:vectorplot}(b). A phase diagram of the global structure in $f$-$\Gamma$ space is displayed in Fig.~\ref{fig:phase_diagram}. As seen in Fig.~\ref{fig:phase_diagram}, the transition occurs at $f\simeq 50$ Hz. The transition is almost independent of $\Gamma$ at least under the current experimental condition. Note that the phase diagram is constructed only based on the experiment with $d=0.8$ mm glass beads. Besides, although we define two types of convective roll structures, this classification is not conclusive. Detail measurements of inner structures of convective rolls are necessary to completely categorize the convective structures. The current classification is still tentative since it is based only on the observation by eye.

\subsection{Analysis of convective velocity}
\label{subsec:v_zmax_calculation}
Next, we analyze the convective velocity. Since the main motivation of this study is to characterize the scaling behavior of the representative convective velocity in a vibrated granular matter, we are going to focus on the toroidal-roll (axisymmetric) state. In the single-roll state, convective velocity is spatially heterogeneous in horizontal direction as shown in Fig.~\ref{fig:vectorplot}(a). As mentioned above, the observable velocity field depends on the camera angle since the rolling direction is stochastically determined and cannot be controlled in the single-roll state. Such a state is inappropriate to extract a representative velocity of the system. By contrast, the toroidal-roll state is rather homogeneous everywhere in $x$ direction of the side wall as shown in Figs.~\ref{fig:vector_and_vertical_plot}(a) and \ref{fig:vectorplot}(b). Furthermore, $v_x$ is always almost zero at any $z$ (Fig.~\ref{fig:vector_and_vertical_plot}(c)), and $v_{z}$ is basically a simple decreasing function of $z$ (Fig.~\ref{fig:vector_and_vertical_plot}(b)) in the toroidal-roll state. \citet{Garcimartin2002} found a qualitatively similar velocity profile in the vibrated granular convection. They also used the maximum velocity to characterize the convective velocity. Therefore, we use the same strategy in the analysis of the representative convective velocity. Namely, we employ the maximum value of the vertical component of velocity, $v_{\rm zmax}$, as a representative convective velocity to characterize the velocity field.

\begin{figure}
\begin{center}
{\includegraphics{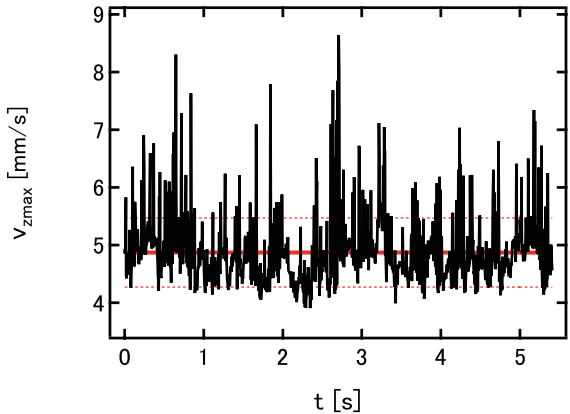}}
\caption{Temporal variation of $v_{\rm zmax}$ at $\Gamma=4$, $f=200$ Hz, $d=0.8$ mm, $H=50$ mm, and $R=37.5$ mm. Intermittent spikes can be observed as well as random fluctuation around the mean value. A horizontal solid line indicates the average value ($v_{\rm zmax}=4.9$ mm/s) and dotted lines show standard deviation levels ($\pm 0.6$ mm/s).}
\label{fig:time_changing}
\end{center}
\end{figure}

Temporal homogeneity of $v_z$ is also examined. In the toroidal-roll state, its spatial structure of the convective velocity is symmetric and simple as discussed above. Then, how about the temporal structure? In Fig.~\ref{fig:time_changing}, a typical time series data of $v_{\rm zmax}(t)$ ($v_{\rm zmax}$ at each time $t$) is shown. While $v_{\rm zmax}$ basically looks more or less steady, a sort of intermittent velocity spikes can be observed in Fig.~\ref{fig:time_changing}. This slight intermittency might come from the inherent complexity of the granular flow due to the frictional effect. 

\begin{figure}
\begin{center}
{\includegraphics{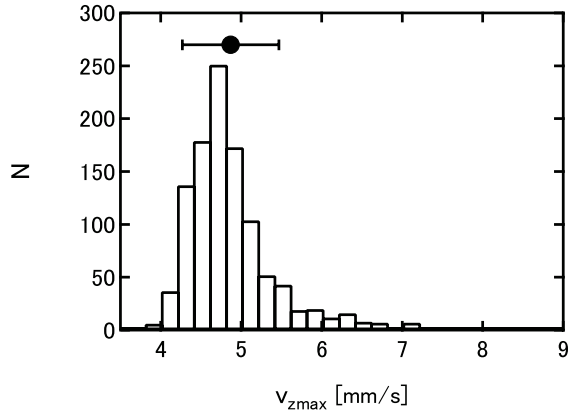}}
\caption{Histogram of $v_{\rm zmax}$ computed from the data in Fig.~\ref{fig:time_changing}. Width of each bin is $0.2$ mm/s. One can confirm the peak of $N$ around $4.7$ mm/s and the asymmetric tail in large velocity regime. Above the histogram, mean and standard deviation are shown as a filled circle and an error bar, respectively ($4.9 \pm 0.6$ mm/s). The corresponding standard error is $\pm 0.02$ mm/s which is smaller than the size of the circle mark.}
\label{fig:histogram}
\end{center}
\end{figure}

The distribution of velocities $N(v_{\rm zmax})$ obtained from the data of Fig.~\ref{fig:time_changing} is displayed in Fig.~\ref{fig:histogram}. Here, $N(v_{\rm zmax})$ indicates the number in the range of $(v_{\rm zmax}, v_{\rm zmax}+dv_{\rm zmax})$ and $dv_{\rm zmax}=0.2$ mm/s is the bin size. The distribution has a clear peak around $v_{\rm zmax} \simeq 4.7$ mm/s. However, an asymmetric tail can be observed at large $v_{\rm zmax}$ regime. Whereas this statistical property of the convective velocity field could contain rich physics, here we neglect the asymmetric structure for simplicity's sake. In Fig.~\ref{fig:histogram}, the mean value and standard deviation of the entire data set are shown as a filled circle and an error bar shown above the histogram. More detail analyses of $v_{\rm zmax}(t)$ by its distribution and temporal correlation etc. are interesting future problems. 

\subsection{Scaling analysis of the convective velocity}
\label{subsec:scaling}
\begin{figure}
\begin{center} 
{\includegraphics{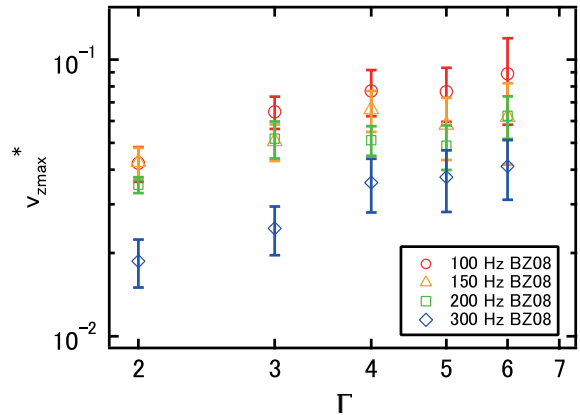}}
\caption{Scaling of $v_{\rm zmax}^{\ast}(=v_{\rm zmax}/\sqrt{gd})$ by $\Gamma$. Each open symbol shows the data of glass beads bed of $d=0.8$ mm, $H=50$ mm, and $R=37.5$ mm. While a systematic correlation between $v_{\rm zmax}^{\ast}$ and $\Gamma$ can be observed, a universal form cannot be obtained solely by $\Gamma$ scaling.}
\label{fig:G_scaling}
\end{center}
\end{figure}

Finally, a scaling of the granular convective velocity is derived. As a first step of the scaling, we focus on the effect of vibration-conditions ($f$ and $\Gamma$) by fixing the geometric conditions: $d=0.8$ mm, $H=50$ mm, and $R=37.5$ mm. As mentioned above, we regard $v_{\rm zmax}$ as a representative granular convective velocity. More precisely, we use the maximum value of $v_z(z)$ profile (e.g., Fig.~\ref{fig:vector_and_vertical_plot}(b)) as $v_{\rm zmax}$. In order to discuss the scale-free nature, a relation among dimensionless parameters should be analyzed. Thus $v_{\rm zmax}$ is normalized to the characteristic velocity created by gravity as,
\begin{equation}
v_{\rm zmax}^{\ast}=v_{\rm zmax}/\sqrt{gd}.
\label{eq:vzmax_star}
\end{equation} 
In Fig.~\ref{fig:G_scaling}, $v_{\rm zmax}^{\ast}$ vs. $\Gamma$ for various $f$ is shown. One can confirm that all the $v_{\rm zmax}^{\ast}$ data are similar increasing functions of $\Gamma$. However, these specific values depend not only on $\Gamma$ but also on $f$. Therefore, it is hard to obtain a universal scale-free relation from this plot. In other words, sole $\Gamma$ cannot characterize the granular convective velocity very well. While \citet{Garcimartin2002} used $\Gamma$ and the normalized bed height to collapse the convective velocity data, here we consider a different quantity to deduce a simple power-law form for the granular convective velocity. Note that the specific functional form of the granular convective velocity was not obtained by \citet{Garcimartin2002}.  

\begin{figure}
\begin{center}
{\includegraphics{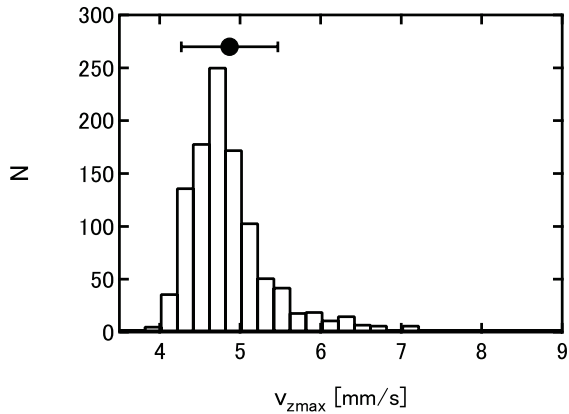}}
\caption{Scaling of $v_{\rm zmax}^{\ast}$ by $S$. Colors and markers used are identical to those in Fig.~\ref{fig:G_scaling}. The dashed line represents the power law fit (Eqs.~(\ref{eq:fitting})). The slope (power) of the dashed line is computed as $\alpha=0.31$. The solid line corresponds to $\alpha=0.47$ which is obtained by the least square search for all the data including various system size parameters (Fig.~\ref{fig:SL_scaling}).
}
\label{fig:S_scaling}
\end{center}
\end{figure}

We find that the shaking parameter $S$ is a relevant dimensionless parameter to scale $v_{\rm zmax}^{\ast}$. Corresponding scaling plot is shown in Fig.~\ref{fig:S_scaling}. All of the $v_{\rm zmax}^{\ast}$ data collapse to a power-law form,  
\begin{equation}
v_{\rm zmax}^{\ast} \sim S^{\alpha},
\label{eq:fitting}
\end{equation}
where $\alpha=0.31$ is a characteristic exponent obtained by the data fitting shown as a dashed line in Fig.~\ref{fig:S_scaling}. The solid line corresponds to $\alpha=0.47$ which is determined by the scaling with various system size data (Fig.~\ref{fig:SL_scaling}). As mentioned before, $S$ represents the balance between the vibrational and gravitational velocities. Thus it is rather natural that another velocity balance (convective velocity vs. gravitational velocity; $v_{\rm zmax}/\sqrt{gd}$) is scaled by $S$. 

The range of $S$ we used in this study is slightly different from that in the previous studies. In \citet{Eshuis2010}, $S$ was used to characterize the onset of convection in the strongly shaken shallow granular bed. In this study, we experimentally confirm that $S$ is also useful to the scaling analysis of the convective velocity in a weakly shaken thick granular bed. The range of $S$ for our experiment is from $O(10^{-2})$ to $O(10^0)$. Although the order of this regime is about one order of magnitude smaller than that in the experiment of~\citet{Eshuis2007} ($S\simeq O(10^1)$), the current experimental result is also well explained by $S$. 

\begin{figure}
\begin{center}
{\includegraphics{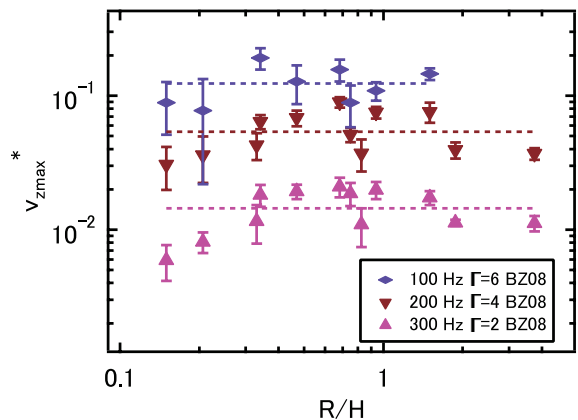}}
\caption{Dependence of $v_{\rm zmax}^{\ast}$ to the aspect ratio of the shaken granular bed, $R/H$. Colors and shapes of markers indicate vibrational conditions as denoted in the legend. All the data points are originated from the experiments with glass beads of $d=0.8$ mm (BZ08). We remove the data of single-roll state from this plot. Although the level of $v_{\rm zmax}^{\ast}$ depends on vibrational conditions, it is roughly independent of $R/H$. Dashed lines are the average levels for each condition.}
\label{fig:aspect_ratio}
\end{center}
\end{figure}

\begin{figure*}
\begin{center}
{\includegraphics{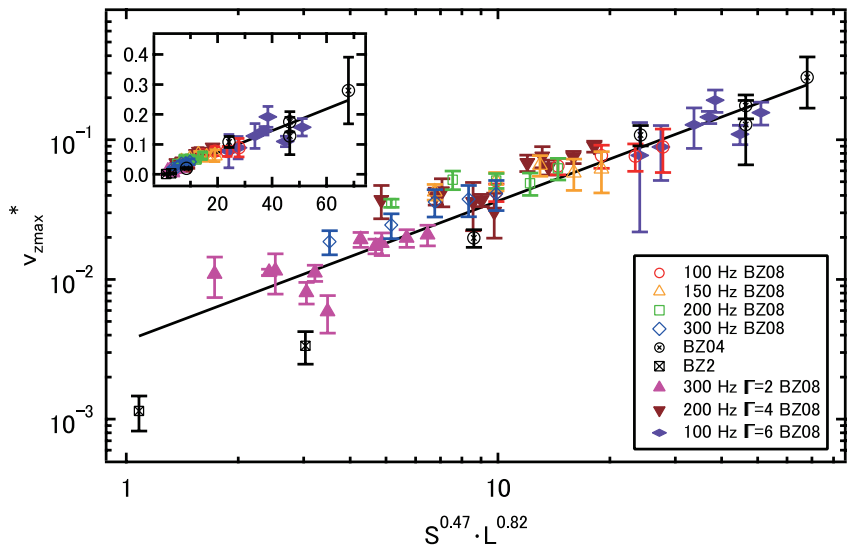}}
\caption{Scaling of $v_{\rm zmax}^{\ast}$ by $S$ and $L$, $v_{\rm zmax}^{\ast} = 3.6 \times 10^{-3}S^{0.47}L^{0.82}$. Colors and markers used are identical to those in Figs.~\ref{fig:S_scaling} and \ref{fig:aspect_ratio}. In addition, black open circles and black open squares correspond to the results with glass beads of BZ04 ($d=0.4$ mm) and BZ2 ($d=2$ mm), respectively. A solid line represents the weighted power-law fit (Eq.~(\ref{eq:SL_scaling})). The data are in good agreement with the scaling. Two data points deviating from the fitting at small scaling variable regime come from the partial crystallization of large beads (BZ2). Inset shows the linear-linear plot of the identical data.}
\label{fig:SL_scaling}
\end{center}
\end{figure*}

As a next step of the scaling, we investigate the system size dependence of $v_{\rm zmax}^{\ast}$ by the varying geometric parameters: bed height $H$ and its radius $R$. In Fig.~\ref{fig:aspect_ratio}, the data of $v_{\rm zmax}^{\ast}$ vs.~$R/H$ for various $\Gamma$ and $f$ are shown. Non-axisymmetric convection patterns such as a single-roll state are sometimes found in the experiment at some $R/H$. Due to the inhomogeneity, the representative velocities in these states show extremely large dispersion. Since we restrict ourselves within the analysis of representative velocity in the homogeneous convection, these data are excluded from the analysis. As shown in Fig.~\ref{fig:aspect_ratio}, we find that $v_{\rm zmax}^{\ast}$ is roughly independent of $R/H$ although its specific value clearly depends on vibrational conditions ($\Gamma$ and $f$). This simple relation $v_{\rm zmax}^{\ast} \sim (R/H)^0$ suggests that $v_{\rm zmax}^{\ast}$ can be scaled equivalently both by $R$ and $H$. Namely, $v_{\rm zmax}^{\ast}$ is the increasing function both of $R$ and $H$. Then we can reduce the number of geometric parameters, i.e., $R$ and $H$ can be unified into a single parameter. In specific, a dimensionless parameter for the system size $L$ can be introduced by using grains diameter $d$ as, 
\begin{equation}
L = \frac{\sqrt{RH}}{d}.
\label{eq:L}
\end{equation}

To prove the validity of the dimensionless parameter $L$, $d$ should be varied as well as $R$ and $H$. Therefore, we also examine the $d$ dependence of $v_{\rm zmax}^{*}$. For this purpose, the aspect ratio is fixed ($R=37.5$ mm and $H=50$ mm) and glass beads of $d=0.4$ mm (BZ04) and $d=2$ mm (BZ2) are used in the experiment. While $v_{\rm zmax}^{\ast}$ of BZ04 is in the same order as the BZ08 case, $v_{\rm zmax}^{\ast}$ of BZ2 is about one order of magnitude smaller than that of BZ08, under the same vibrational condition. This is due to the effect of crystallization which prevents the convective motion. Besides, the solidification (complete crystallization) of glass beads bed occurs when BZ2 is shaken by $\Gamma\ge3$. The crystallization is unavoidable for the larger grains as long as we use nearly mono-disperse spherical grains. Whereas the number of data points is not enough to discuss the specific scaling form, we qualitatively confirm that the convective velocity is a decreasing function of $d$.  

Altogether, we can assume a simple scaling relation among these dimensionless parameters to compile all the data. The assumed scaling form is written as, 
\begin{equation}
v_{\rm zmax}^{\ast} \sim S^{\alpha}L^{\beta}.
\label{eq:SL_scaling}
\end{equation}
The values of $\alpha$ and $\beta$ are computed by searching the least square point in the range of $0 \le \alpha \le 2$ and $0 \le \beta \le 2$. The least square fitting in Fig.~(\ref{fig:SL_scaling}) is computed with weighting of the standard deviation (velocity scattering degree). The obtained values are $\alpha=0.47$ and $\beta=0.82$. And the computed numerical prefactor is $3.6 \times 10^{-3}$. The scaling result for all the experimentally obtained data is shown in Fig.~\ref{fig:SL_scaling}. As can be seen, the data are basically well scaled by the variable $S^{\alpha}L^{\beta}$. The corresponding scaling line ($v_{\rm zmax}^{\ast} \sim S^{0.47}$) is also plotted in Fig.~\ref{fig:S_scaling} as a solid line. The value of $\alpha$ is slightly improved from the fixed system size case (Fig.~(\ref{fig:S_scaling})). In Fig.~\ref{fig:SL_scaling}, two data points deviate significantly from the scaling. This deviation comes from the crystallization effect of large glass beads (BZ2). The inset of Fig.~\ref{fig:SL_scaling} shows a linear plot of the same data. This scaling relation is the main result of this experimental investigation.

\section{Discussion}
\label{sec:Discussion}
\subsection{Physical meaning and limitations of the scaling}
\label{subsec:physical_meaning}
There are some missing parameters in the scaling of Eq.~(\ref{eq:SL_scaling}). For example, density is not involved in it. That is, the scaling is a kind of kinematic one. The density dependence has to be clarified to understand the dynamic scaling including the effect of inertia~\citep{Hubbert1937}. Perhaps, this would not be so critical because the density of glass beads ($2.5 \times 10^3$ kg/m$^3$) is close to (at least in the same order with) the real regolith density (e.g., $3.4 \times 10^3$ kg/m$^3$ for Itokawa~\citep{Tsuchiyama2011}). Another example of the missing parameters is the cohesive force among grains. Although the gravitational force is very small on the surface of small asteroids, the cohesive effect would instead play a crucial role~\citep{Scheeres2010}. It is hard to check this effect by the current experiment since we do not vary the cohesiveness of grains. In the current experiment, the cohesive force is negligible since it is much smaller than the gravitational force. We assume that interstitial air and electrostatic effects do not play essential roles, i.e., they are much smaller than gravity effect. Indeed, the air effect is negligible as long as we use $d \geq 0.4$ mm grains (see e.g., Supplementary information of \citet{Katsuragi2007}). We also think that the gravitational force is more dominant than the electrostatic force for the bulk convective motion of glass beads. In fact, while we saw a few glass beads sticking on the sidewall due to the electrostatic effect, those were really few. The electrostatic effect might slightly affect the convective motion. However, we consider it cannot be a dominant effect.

A much more serious problem is the absence of frictional properties in the scaling. The frictional properties are a possible origin of intermittent velocity fluctuation shown in Fig.~\ref{fig:time_changing}. In addition, it might affect the representative value of convective velocity. Frictional properties relate to the shape of constitutive grains. Obviously, actual regolith grains' shapes are irregular and rough. By a preliminary experiment, we actually confirm that $v_{\rm zmax}$ significantly reduces when the rough-shaped sand grains are used. On the contrary, some previous studies have reported that the friction can enhance the convective velocity~\citep{Clement1992,Elperin1997}. Moreover, the grain-wall friction is different from grain-grain friction. The grain-wall friction could be a reason of the onset of granular convection in this experiment. The detailed study for the effect of friction is a very important future problem.

The origin of granular convection in the current experiment is plausibly the effect of wall. As already mentioned, the structure of the convective roll changes spontaneously when $f$ or $R$ are varied. This implies that there is an inherent length scale in the granular convection. However, the wall effect to the velocity scaling is still significant even in split (doubled) roll state since we observe the grains' motion on the wall. Moreover, it has been shown that experimental conditions including the container's shape and interstitial air influence the convective roll structure~\citep{Knight1993,Aoki1996,Pak1995}. In these previous studies, the interaction between grains and wall seems to be a key factor. Nevertheless, astronomical objects obviously maintain their own shapes without any wall. It is hard to eliminate the container wall effect in usual granular experiments even in microgravity environment. Rather, highly heterogenous structure of small asteroids might work as an effective wall which triggers the granular convection. Put another way, there could be a lot of effective boundaries and/or gradients of frictional properties in actual small asteroids. Thus it is not quite clear whether the granular convection occurs globally or locally in small asteroids.

The structure of the velocity profile $v_z(z)$ is informative both for the granular physics and planetary science. In the previous studies of granular convection~\citep{Taguchi1992,Ehrichs1995}, convective velocity decays rapidly (exponentially) with depth. And the deep region of the vibrated granular matter is almost at rest. If this peculiar regime called {\it frozen zone} presents in the real asteroids' convection, the regolith migration occurs only within a shallow region. In the current experiment, most of velocity profiles do not show the clear frozen zone. The convective velocity decays calmly as typically shown in Fig.~\ref{fig:vector_and_vertical_plot}(b). Thus the convective velocity is simply scaled by $L$ with the nontrivial exponent $\beta = 0.82$. If the frozen zone is clearly observed, the convective velocity would be saturated at a certain $H$, and it cannot be simply scaled by a single exponent. Actually, the frozen zone can be partially observed also in the current experiment under small $\Gamma$, large $f$, and large $H$ conditions. Since we are interested in the principal flow velocity of the granular convection, we have focused only on the axisymmetric convective state and its maximum velocity $v_{\rm zmax}$, in this study. Detailed analysis of $v_z(z)$ profile is a next important step to reveal the granular convection dynamics that relates to the depth of regolith convection occurring in small astronomical objects. 

\subsection{Dimensional analysis of the granular convective velocity}
\label{subsec:dimensional_analysis}
So far, the scaling form for the granular convective velocity has been discussed on the basis of systematically performed experimental results. In this subsection, we discuss the implication of the obtained scaling law. The scaling of Eq.~(\ref{eq:SL_scaling}) can be rewritten by considering its physical meaning. Then the scaling is compared with a recent experimental result of two-dimensional granular convection.

Equation~(\ref{eq:SL_scaling}) can be transformed into a following form,
\begin{equation}
v_{\rm zmax} \sim \left ( \sqrt{gd} \right )^{1-2\alpha} \left (2 \pi A_0 f \right )^{2\alpha} \left (\frac{\sqrt{RH}}{d} \right )^{\beta}.
\label{eq:scaling1}
\end{equation}
This form is more understandable than Eq.~(\ref{eq:SL_scaling}). The representative velocity $v_{\rm zmax}$ is expressed by a power law product of two characteristic velocities: gravitational $\sqrt{gd}$ and vibrational $2\pi A_0 f$. In addition, the system size dependence is scaled by the last factor $\sqrt{RH}/d$. 

The scaling of Eq.~(\ref{eq:scaling1}) is qualitatively consistent with a recent experiment performed by~\citet{Hejmady2012}. They investigated the convective-driven BNE and found that the convective velocity linearly depends on $2\pi A_0 f$. To better explain the experimental data, the use of $2\pi A_0 f$ was much better than $\Gamma$. According to Eq.~(\ref{eq:scaling1}), convective velocity depends on vibrational velocity as, $v_{\rm zmax} \propto (2\pi A_0 f)^{0.94}$. These two experimental results show a good agreement. However, they reported the presence of critical vibrational velocity below which the granular convection was not induced. This critical vibrational velocity probably corresponds to the onset criterion. In the current analysis, however, we do not clearly confirm this offset in the scaling (see Fig.~\ref{fig:SL_scaling}). Instead, we varied $S$ over two orders of magnitude to obtain the power-law relation. The range of $S$ and other parameters were relatively narrow in the study of \citet{Hejmady2012} since they are not interested in the scale-free form. Thus they reported the linear relation between the convective velocity and the vibrational velocity. On the other hand, we obtain the scaling (power law) relation between the convective velocity, the vibrational velocity, and the gravitational velocity. 

It should be noted that the system size dependence of the scaling is bounded. According to Eq.~(\ref{eq:scaling1}), the value of $v_{\rm zmax}$ can be very large when $L=\sqrt{RH}/d$ becomes large. This condition corresponds to a large $\sqrt{RH}$ or a small $d$ case. At the limit of $\sqrt{RH} \to \infty$ or $d \to 0$, $v_{\rm zmax}$ diverges. However, such a divergence does not occur in the actual granular convection. The convective roll size cannot grow to the infinitely large size. In fact, we experimentally observed the split of the convective roll (double toroidal-roll) in a very shallow layer, although such a split rarely happened. This preliminary observation indicates that there might be an intrinsic length scale which determines the maximum convective roll size. Since the current experimental system size is limited, it is hard to systematically reveal the effect of this intrinsic length scale. At a small $d$ regime, on the other hand, the cohesive force among grains becomes dominant instead of the gravitational force. The current scaling form cannot be applied to such a situation. The detailed investigation on the size-dependent scaling form for more widely spreading size regime is an important future problem. To investigate the small $d$ regime, the experiment under the vacuum condition is also necessary.

\subsection{Gravity effect to the granular convective velocity}
\label{subsec:gravity_effect}
Finally, the experimentally obtained scaling is cast into a form in which the gravity effect can be directly evaluated. Since the interpretation of scaling is not unique in general, we have to take a special care for its understanding. Whereas the discussion in this subsection is speculative, the result is consistent with other related works. Moreover, the obtained scaling includes various parameters such as vibration frequency and system size. Thus the form is very efficient to estimate the timescale of regolith convection occurring on the surface of small asteroid.

Recently, some granular flow experiments were performed under microgravity conditions created by parabolic flights~\citep{Murdoch2013a,Murdoch2013b,Murdoch2013c,Guttler2013}. \citet{Murdoch2013a,Murdoch2013b,Murdoch2013c} used a Taylor-Couette geometry and measured both sheared and convective flows induced by the shearing. Particularly, \citet{Murdoch2013a} reported the effect of gravity to the sheared granular convection. They claimed that the gravity plays an essential role to make a stiff grains network in a bulk granular matter. Such a grains network is necessary to transform a shear stress into a driving force of the convective motion. Therefore, the granular convective velocity under the microgravity became almost zero. Moreover, the convective velocity under the high-gravity condition became large~\citep{Murdoch2013a}. \citet{Guttler2013} measured the rising velocity of intruders in convective-driven BNE under the reduced gravity. They found that the rising velocity is almost linearly scaled by the gravity. 

To compare the current result with above-mentioned microgravity experiments, here we evaluate the gravity influence to the convective velocity based on Eq.~(\ref{eq:scaling1}). At a glance of Eq.~(\ref{eq:scaling1}), the convective velocity is scaled as $v_{\rm zmax} \propto g^{0.03} ((1-2\alpha)/2=0.03)$. This scaling implicitly assumes the fixed $2\pi A_0 f$ under different $g$ environment. This assumption is not practical to estimate the granular convective velocity induced under the different $g$. For instance, under the microgravity environment, very small vibrational acceleration is enough to satisfy the onset criterion of granular convection ($\Gamma > 1$). However, if $2\pi A_0 f$ is fixed to consider the granular convective velocity under the microgravity environment, the corresponding $\Gamma$ becomes extremely large. It is not a realistic assumption. Instead, the value of $\Gamma$ should be fixed to estimate the convective velocity by means of similarity law. In fact, fixed $\Gamma$ was used to discuss the possibility of convection on the surface of Itokawa~\citep{Miyamoto2007}. \citet{Guttler2013} also fixed $\Gamma$ to discuss $g$ dependence. Therefore, the typical vibrational velocity should be written as $\Gamma g/2\pi f$ instead of $2\pi A_0 f$. Then, Eq.~(\ref{eq:scaling1}) can be modified into the following form, 
\begin{equation}
v_{\rm zmax} \sim \left ( \sqrt{gd} \right )^{1-2\alpha} \left (\frac{\Gamma g}{2 \pi f} \right )^{2\alpha} \left (\frac{\sqrt{RH}}{d} \right )^{\beta}.
\label{eq:scaling2}
\end{equation}
According to Eq.~(\ref{eq:scaling2}), $v_{\rm zmax}$ depends on $g$ as, $v_{\rm zmax} \propto g^{(2\alpha +1)/2}$. The value $(2\alpha +1)/2=0.97$ is close to unity. This result is qualitatively similar to \citet{Murdoch2013a,Murdoch2013b,Murdoch2013c} and \citet{Guttler2013}. However, note that we do not directly vary the gravitational acceleration. The $g$ dependence is indirectly deduced from the scaling. Besides, the convective motion of~\citet{Murdoch2013a,Murdoch2013b,Murdoch2013c} was induced by shear in a Taylor-Couette cell. Thus the similarity is rather qualitative. 

Using Eq.~(\ref{eq:scaling2}) and a model of seismic shaking induced by impacts~\citep{Richardson2004, Richardson2005,Miyamoto2007}, we are able to estimate the convective velocity for various astronomical objects. To do that, we have to be careful in the estimate of $\sqrt{RH}/d$. Whereas its value strongly affects the estimate of $v_{\rm zmax}$, it is still very uncertain. Then it might be possible to evaluate the surface age (resurfacing timescale) of the microgravity asteroid by considering the population of impactors. Such an estimate for Itokawa or Eros is an interesting application of this scaling. This topic is a part of ongoing investigation.

\section{Conclusions}
\label{sec:Conclusions}
An experimental study of the granular convection was performed with a vertically shaken glass beads bed. The maximum convective velocity $v_{\rm zmax}$ was measured by PIV method. Although the $v_{\rm zmax}$ was measured on the wall, we assume that it corresponds to a representative value of the convective velocity. The normalized convective velocity $v_{\rm zmax}^{\ast}=v_{\rm zmax}/\sqrt{gd}$ was scaled by the shaking parameter $S=(2\pi A_0 f)^2/gd$ and the dimensionless system size $L=\sqrt{RH}/d$. The obtained scaling (Eq~(\ref{eq:scaling1})) indicates that the representative convective velocity can be written as a certain power-law product of the vibrational velocity $2\pi A_0 f$, the gravitational velocity $\sqrt{gd}$, and the dimensionless system size $L$. This experimentally obtained scaling law is qualitatively consistent with other previous studies of granular convection or BNE. Using the scaling law, one can roughly estimate the convective velocity for various vibrational conditions, in principle. Since the gravitational acceleration and frictional properties are not varied in this experiment, applicability of the current result toward actual regolith migration might be still limited. Systematic further studies have to be carried out to unveil the regolith migration dynamics owing to the granular convection.  

\section*{Acknowledgments}
We would like to thank to S. Watanabe, H. Kumagai, S. Sirono, and T. Morota for fruitful discussions and suggestions. This research has been partly supported by JSPS KAKENHI Grant Number 23654134 and Nagoya University Program for Leading Graduate Schools (Leadership Development Program for Space Exploration and Research).


\begin{thebibliography}{39}
\expandafter\ifx\csname natexlab\endcsname\relax\def\natexlab#1{#1}\fi
\providecommand{\url}[1]{\texttt{#1}}
\providecommand{\href}[2]{#2}
\providecommand{\path}[1]{#1}
\providecommand{\DOIprefix}{doi:}
\providecommand{\ArXivprefix}{arXiv:}
\providecommand{\URLprefix}{URL: }
\providecommand{\Pubmedprefix}{pmid:}
\providecommand{\doi}[1]{\href{http://dx.doi.org/#1}{\path{#1}}}
\providecommand{\Pubmed}[1]{\href{pmid:#1}{\path{#1}}}
\providecommand{\bibinfo}[2]{#2}
\ifx\xfnm\relax \def\xfnm[#1]{\unskip,\space#1}\fi
\bibitem[{Abe et~al.(2006)Abe, Mukai, Hirata, Barnouin-jha, Cheng, Demura,
  Gaskell, Kubota, Matsuoka, Mizuno, Nakamura, Sheeres and Yoshikawa}]{Abe2006}
\bibinfo{author}{Abe, S.}, \bibinfo{author}{Mukai, N.},
  \bibinfo{author}{Hirata, N.}, \bibinfo{author}{Barnouin-jha, O.S.},
  \bibinfo{author}{Cheng, A.F.}, \bibinfo{author}{Demura, H.},
  \bibinfo{author}{Gaskell, R.W.}, \bibinfo{author}{Kubota, T.},
  \bibinfo{author}{Matsuoka, M.}, \bibinfo{author}{Mizuno, T.},
  \bibinfo{author}{Nakamura, R.}, \bibinfo{author}{Sheeres, D.J.},
  \bibinfo{author}{Yoshikawa, M.}, \bibinfo{year}{2006}.
\newblock \bibinfo{title}{Mass and local topography measurements of itokawa by
  hayabusa}.
\newblock \bibinfo{journal}{Science} \bibinfo{volume}{312},
  \bibinfo{pages}{1344--1347}.
\bibitem[{Aoki et~al.(1996)Aoki, Akiyama, Maki and Watanabe}]{Aoki1996}
\bibinfo{author}{Aoki, K.M.}, \bibinfo{author}{Akiyama, T.},
  \bibinfo{author}{Maki, Y.}, \bibinfo{author}{Watanabe, T.},
  \bibinfo{year}{1996}.
\newblock \bibinfo{title}{Convective roll patterns in vertically vibrated beds
  of granules}.
\newblock \bibinfo{journal}{Phys. Rev. E} \bibinfo{volume}{54},
  \bibinfo{pages}{874--883}.
\bibitem[{Bokkers et~al.(2004)Bokkers, van Sint~Annaland and
  Kuipers}]{Bokkers2004}
\bibinfo{author}{Bokkers, G.}, \bibinfo{author}{van Sint~Annaland, M.},
  \bibinfo{author}{Kuipers, J.}, \bibinfo{year}{2004}.
\newblock \bibinfo{title}{Mixing and segregation in a bidisperse gas - solid
  fluidised bed: a numerical and experimental study}.
\newblock \bibinfo{journal}{Powder Technology} \bibinfo{volume}{140},
  \bibinfo{pages}{176 -- 186}.
\bibitem[{Bottke et~al.(2002)Bottke, Cellino, Paolicchi and
  Binzel}]{AsteroidsIII}
\bibinfo{editor}{Bottke, W.F.}, \bibinfo{editor}{Cellino, A.},
  \bibinfo{editor}{Paolicchi, P.}, \bibinfo{editor}{Binzel, R.P.} (Eds.),
  \bibinfo{year}{2002}.
\newblock \bibinfo{title}{Asteroids III}.
\newblock \bibinfo{publisher}{University of Arizona Press},
  \bibinfo{address}{Tucson}.
\bibitem[{Cl\'ement et~al.(1992)Cl\'ement, Duran and Rajchenbach}]{Clement1992}
\bibinfo{author}{Cl\'ement, E.}, \bibinfo{author}{Duran, J.},
  \bibinfo{author}{Rajchenbach, J.}, \bibinfo{year}{1992}.
\newblock \bibinfo{title}{Experimental study of heaping in a two-dimensional
  ``sand pile''}.
\newblock \bibinfo{journal}{Phys. Rev. Lett.} \bibinfo{volume}{69},
  \bibinfo{pages}{1189--1192}.
\bibitem[{Ehrichs et~al.(1995)Ehrichs, Jaeger, Karczmar, Knight, Kuperman and
  Nagel}]{Ehrichs1995}
\bibinfo{author}{Ehrichs, E.E.}, \bibinfo{author}{Jaeger, H.M.},
  \bibinfo{author}{Karczmar, G.S.}, \bibinfo{author}{Knight, J.B.},
  \bibinfo{author}{Kuperman, V.Y.}, \bibinfo{author}{Nagel, S.R.},
  \bibinfo{year}{1995}.
\newblock \bibinfo{title}{Granular convection observed by magnetic resonance
  imaging}.
\newblock \bibinfo{journal}{Science} \bibinfo{volume}{267},
  \bibinfo{pages}{1632--1634}.
\bibitem[{Elperin and Golshtein(1997)}]{Elperin1997}
\bibinfo{author}{Elperin, T.}, \bibinfo{author}{Golshtein, E.},
  \bibinfo{year}{1997}.
\newblock \bibinfo{title}{Effects of convection and friction on size
  segregation in vibrated granular beds}.
\newblock \bibinfo{journal}{Physica A: Statistical Mechanics and its
  Applications} \bibinfo{volume}{247}, \bibinfo{pages}{67--78}.
\bibitem[{Eshuis et~al.(2010)Eshuis, van~der Meer, Alam, van Gerner, van~der
  Weele and Lohse}]{Eshuis2010}
\bibinfo{author}{Eshuis, P.}, \bibinfo{author}{van~der Meer, D.},
  \bibinfo{author}{Alam, M.}, \bibinfo{author}{van Gerner, H.J.},
  \bibinfo{author}{van~der Weele, K.}, \bibinfo{author}{Lohse, D.},
  \bibinfo{year}{2010}.
\newblock \bibinfo{title}{Onset of convection in strongly shaken granular
  matter}.
\newblock \bibinfo{journal}{Phys. Rev. Lett.} \bibinfo{volume}{104},
  \bibinfo{pages}{038001}.
\bibitem[{Eshuis et~al.(2007)Eshuis, van~der Weele, van~der Meer, Bos and
  Lohse}]{Eshuis2007}
\bibinfo{author}{Eshuis, P.}, \bibinfo{author}{van~der Weele, K.},
  \bibinfo{author}{van~der Meer, D.}, \bibinfo{author}{Bos, R.},
  \bibinfo{author}{Lohse, D.}, \bibinfo{year}{2007}.
\newblock \bibinfo{title}{Phase diagram of vertically shaken granular matter}.
\newblock \bibinfo{journal}{Phisics of Fluids} \bibinfo{volume}{19},
  \bibinfo{pages}{123301--1}.
\bibitem[{Eshuis et~al.(2005)Eshuis, van~der Weele, van~der Meer and
  Lohse}]{Eshuis2005}
\bibinfo{author}{Eshuis, P.}, \bibinfo{author}{van~der Weele, K.},
  \bibinfo{author}{van~der Meer, D.}, \bibinfo{author}{Lohse, D.},
  \bibinfo{year}{2005}.
\newblock \bibinfo{title}{Granular leidenfrost effect: Experiment and theory of
  floating particle clusters}.
\newblock \bibinfo{journal}{Phys. Rev. Lett.} \bibinfo{volume}{95},
  \bibinfo{pages}{258001}.
\bibitem[{Faraday(1831)}]{Faraday1831}
\bibinfo{author}{Faraday, M.}, \bibinfo{year}{1831}.
\newblock \bibinfo{title}{On a peculiar class of acoustical figures; and on
  certain forms assumed by groups of particles upon vibrating elastic
  surfaces}.
\newblock \bibinfo{journal}{Philosophical Transactions of the Royal Society of
  London} \bibinfo{volume}{52}, \bibinfo{pages}{299}.
\bibitem[{Fujiwara et~al.(2006)Fujiwara, Kawaguchi, Yeomans, Abe, Mukai, Okada,
  Saito, Yano, Yoshikawa, J., Barnouin-Jha, Cheng, Demura, Gaskell, Hirata,
  Ikeda, Kominato, Miyamoto, Nakamura, Sasaki and Uesugi}]{Fujiwara2006}
\bibinfo{author}{Fujiwara, A.}, \bibinfo{author}{Kawaguchi, J.},
  \bibinfo{author}{Yeomans, D.K.}, \bibinfo{author}{Abe, M.},
  \bibinfo{author}{Mukai, T.}, \bibinfo{author}{Okada, T.},
  \bibinfo{author}{Saito, J.}, \bibinfo{author}{Yano, H.},
  \bibinfo{author}{Yoshikawa, M.}, \bibinfo{author}{J., S.D.},
  \bibinfo{author}{Barnouin-Jha, O.}, \bibinfo{author}{Cheng, A.F.},
  \bibinfo{author}{Demura, H.}, \bibinfo{author}{Gaskell, R.},
  \bibinfo{author}{Hirata, N.}, \bibinfo{author}{Ikeda, I.},
  \bibinfo{author}{Kominato, T.}, \bibinfo{author}{Miyamoto, H.~Nakamura,
  A.M.}, \bibinfo{author}{Nakamura, R.}, \bibinfo{author}{Sasaki, S.},
  \bibinfo{author}{Uesugi, K.}, \bibinfo{year}{2006}.
\newblock \bibinfo{title}{The rubble-pile asteroid itokawa as observed by
  hayabusa}.
\newblock \bibinfo{journal}{Science} \bibinfo{volume}{312},
  \bibinfo{pages}{1330--1334}.
\bibitem[{Garcimart\'in et~al.(2002)Garcimart\'in, Maza, Ilquimiche and
  Zuriguel}]{Garcimartin2002}
\bibinfo{author}{Garcimart\'in, A.}, \bibinfo{author}{Maza, D.},
  \bibinfo{author}{Ilquimiche, J.L.}, \bibinfo{author}{Zuriguel, I.},
  \bibinfo{year}{2002}.
\newblock \bibinfo{title}{Convective motion in a vibrated granular layer}.
\newblock \bibinfo{journal}{Phys. Rev. E} \bibinfo{volume}{65},
  \bibinfo{pages}{031303}.
\bibitem[{G\"uttler et~al.(2013)G\"uttler, von Borstel, Schr\"apler and
  Blum}]{Guttler2013}
\bibinfo{author}{G\"uttler, C.}, \bibinfo{author}{von Borstel, I.},
  \bibinfo{author}{Schr\"apler, R.}, \bibinfo{author}{Blum, J.},
  \bibinfo{year}{2013}.
\newblock \bibinfo{title}{Granular convection and the brazil nut effect in
  reduced gravity}.
\newblock \bibinfo{journal}{Phys. Rev. E} \bibinfo{volume}{87},
  \bibinfo{pages}{044201}.
\bibitem[{Hejmady et~al.(2012)Hejmady, Bandyopadhyay, Sabhapandit and
  Dhar}]{Hejmady2012}
\bibinfo{author}{Hejmady, P.}, \bibinfo{author}{Bandyopadhyay, R.},
  \bibinfo{author}{Sabhapandit, S.}, \bibinfo{author}{Dhar, A.},
  \bibinfo{year}{2012}.
\newblock \bibinfo{title}{Scaling behavior in the convection-driven brazil nut
  effect}.
\newblock \bibinfo{journal}{Phys. Rev. E} \bibinfo{volume}{86},
  \bibinfo{pages}{050301}.
\bibitem[{Hirata et~al.(2009)Hirata, Barnouin-Jha, Honda, Nakamura, Miyamoto,
  Sasaki, Demura, Nakamura, Michikami, Gaskell and Saito}]{Hirata2010}
\bibinfo{author}{Hirata, N.}, \bibinfo{author}{Barnouin-Jha, O.S.},
  \bibinfo{author}{Honda, C.}, \bibinfo{author}{Nakamura, R.},
  \bibinfo{author}{Miyamoto, H.}, \bibinfo{author}{Sasaki, S.},
  \bibinfo{author}{Demura, H.}, \bibinfo{author}{Nakamura, A.M.},
  \bibinfo{author}{Michikami, T.}, \bibinfo{author}{Gaskell, R.W.},
  \bibinfo{author}{Saito, J.}, \bibinfo{year}{2009}.
\newblock \bibinfo{title}{A survey of possible impact structures on 25143
  itokawa}.
\newblock \bibinfo{journal}{Icarus} \bibinfo{volume}{200},
  \bibinfo{pages}{486--502}.
\bibitem[{Hubbert(1937)}]{Hubbert1937}
\bibinfo{author}{Hubbert, M.K.}, \bibinfo{year}{1937}.
\newblock \bibinfo{title}{Theory of scale models as applied to the study of
  geologic structures}.
\newblock \bibinfo{journal}{Bull. Geol. Soc. Amer.} \bibinfo{volume}{48},
  \bibinfo{pages}{1459--1519}.
\bibitem[{Katsuragi and Durian(2007)}]{Katsuragi2007}
\bibinfo{author}{Katsuragi, H.}, \bibinfo{author}{Durian, D.J.},
  \bibinfo{year}{2007}.
\newblock \bibinfo{title}{Unified force law for granular impact cratering}.
\newblock \bibinfo{journal}{Nature Phys.} \bibinfo{volume}{3},
  \bibinfo{pages}{420--423}.
\bibitem[{Knight et~al.(1996)Knight, Ehrichs, Kuperman, Flint, Jaeger and
  Nagel}]{Knight1996}
\bibinfo{author}{Knight, J.B.}, \bibinfo{author}{Ehrichs, E.E.},
  \bibinfo{author}{Kuperman, V.Y.}, \bibinfo{author}{Flint, J.K.},
  \bibinfo{author}{Jaeger, H.M.}, \bibinfo{author}{Nagel, S.R.},
  \bibinfo{year}{1996}.
\newblock \bibinfo{title}{Experimental study of granular convection}.
\newblock \bibinfo{journal}{Phys. Rev. E} \bibinfo{volume}{54},
  \bibinfo{pages}{5726--5738}.
\bibitem[{Knight et~al.(1993)Knight, Jaeger and Nagel}]{Knight1993}
\bibinfo{author}{Knight, J.B.}, \bibinfo{author}{Jaeger, H.M.},
  \bibinfo{author}{Nagel, S.R.}, \bibinfo{year}{1993}.
\newblock \bibinfo{title}{Vibration-induced size separation in granular media:
  The convection connection}.
\newblock \bibinfo{journal}{Phys. Rev. Lett.} \bibinfo{volume}{70},
  \bibinfo{pages}{3728--3731}.
\bibitem[{Luding et~al.(1994)Luding, Cl\'ement, Blumen, Rajchenbach and
  Duran}]{Luding1994}
\bibinfo{author}{Luding, S.}, \bibinfo{author}{Cl\'ement, E.},
  \bibinfo{author}{Blumen, A.}, \bibinfo{author}{Rajchenbach, J.},
  \bibinfo{author}{Duran, J.}, \bibinfo{year}{1994}.
\newblock \bibinfo{title}{Onset of convection in molecular dynamics simulations
  of grains}.
\newblock \bibinfo{journal}{Phys. Rev. E} \bibinfo{volume}{50},
  \bibinfo{pages}{R1762--R1765}.
\bibitem[{Lueptow et~al.(2000)Lueptow, Akonur and Shinbrot}]{Lueptow2000}
\bibinfo{author}{Lueptow, R.M.}, \bibinfo{author}{Akonur, A.},
  \bibinfo{author}{Shinbrot, T.}, \bibinfo{year}{2000}.
\newblock \bibinfo{title}{{PIV} for granular flows}.
\newblock \bibinfo{journal}{Experiments in Fluids.} \bibinfo{volume}{28},
  \bibinfo{pages}{183--186}.
\bibitem[{Michel et~al.(2009)Michel, O'Brien, Abe and Hirata}]{Michel2009}
\bibinfo{author}{Michel, P.}, \bibinfo{author}{O'Brien, D.},
  \bibinfo{author}{Abe, S.}, \bibinfo{author}{Hirata, N.},
  \bibinfo{year}{2009}.
\newblock \bibinfo{title}{Itokawa's cratering record as observed by hayabusa:
  Implications for its age and collisional history}.
\newblock \bibinfo{journal}{Icarus} \bibinfo{volume}{200},
  \bibinfo{pages}{503--513}.
\bibitem[{Miyamoto et~al.(2007)Miyamoto, Yano, Scheeres, Abe, Barnouin-Jha,
  Cheng, Demura, Gaskell, Hirata, Ishiguro, Michikami, Nakamura, Nakamura,
  Saito and Sasaki}]{Miyamoto2007}
\bibinfo{author}{Miyamoto, H.}, \bibinfo{author}{Yano, H.},
  \bibinfo{author}{Scheeres, J.D.}, \bibinfo{author}{Abe, S.},
  \bibinfo{author}{Barnouin-Jha, O.}, \bibinfo{author}{Cheng, A.F.},
  \bibinfo{author}{Demura, H.}, \bibinfo{author}{Gaskell, R.W.},
  \bibinfo{author}{Hirata, N.}, \bibinfo{author}{Ishiguro, R.},
  \bibinfo{author}{Michikami, T.}, \bibinfo{author}{Nakamura, A.M.},
  \bibinfo{author}{Nakamura, R.}, \bibinfo{author}{Saito, J.},
  \bibinfo{author}{Sasaki, A.}, \bibinfo{year}{2007}.
\newblock \bibinfo{title}{Regolith migration and sorting on asteroid itokawa}.
\newblock \bibinfo{journal}{Science} \bibinfo{volume}{316},
  \bibinfo{pages}{1011--1014}.
\bibitem[{Murdoch et~al.(2013a)Murdoch, Rozitis, Green, Michel, de~Lophem and
  Losert}]{Murdoch2013c}
\bibinfo{author}{Murdoch, N.}, \bibinfo{author}{Rozitis, B.},
  \bibinfo{author}{Green, S.F.}, \bibinfo{author}{Michel, P.},
  \bibinfo{author}{de~Lophem, T.L.}, \bibinfo{author}{Losert, W.},
  \bibinfo{year}{2013}a.
\newblock \bibinfo{title}{Simulating regoliths in microgravity}.
\newblock \bibinfo{journal}{Monthly Notices of the Royal Astronomical Society}
  \bibinfo{volume}{433}, \bibinfo{pages}{506--514}.
\bibitem[{Murdoch et~al.(2013b)Murdoch, Rozitis, Nordstrom, Green, Michel,
  de~Lophem and Losert}]{Murdoch2013a}
\bibinfo{author}{Murdoch, N.}, \bibinfo{author}{Rozitis, B.},
  \bibinfo{author}{Nordstrom, K.}, \bibinfo{author}{Green, S.F.},
  \bibinfo{author}{Michel, P.}, \bibinfo{author}{de~Lophem, T.L.},
  \bibinfo{author}{Losert, W.}, \bibinfo{year}{2013}b.
\newblock \bibinfo{title}{Granular convection in microgravity}.
\newblock \bibinfo{journal}{Phys. Rev. Lett.} \bibinfo{volume}{110},
  \bibinfo{pages}{018307}.
\bibitem[{Murdoch et~al.(2013c)Murdoch, Rozitis, Nordstrom, Green, Michel,
  de~Lophem and Losert}]{Murdoch2013b}
\bibinfo{author}{Murdoch, N.}, \bibinfo{author}{Rozitis, B.},
  \bibinfo{author}{Nordstrom, K.}, \bibinfo{author}{Green, S.F.},
  \bibinfo{author}{Michel, P.}, \bibinfo{author}{de~Lophem, T.L.},
  \bibinfo{author}{Losert, W.}, \bibinfo{year}{2013}c.
\newblock \bibinfo{title}{Granular shear flow in varying gravitational
  environments}.
\newblock \bibinfo{journal}{Granular Matter} \bibinfo{volume}{15},
  \bibinfo{pages}{129--137}.
\bibitem[{Nagao et~al.(2011)Nagao, Okazaki, Nakamura, Miura, Osawa, Bajo,
  Matsuda, Ebihara, Noguchi, Tsuchiyama, Yurimoto, Zolensky, Uesugi, Shirai,
  Abe, Yada, Ishibashi, Fujimura, Mukai, Ueno, Okada and Yoshikawa}]{Nagao2011}
\bibinfo{author}{Nagao, K.}, \bibinfo{author}{Okazaki, R.},
  \bibinfo{author}{Nakamura, T.}, \bibinfo{author}{Miura, N.Y.},
  \bibinfo{author}{Osawa, T.}, \bibinfo{author}{Bajo, K.},
  \bibinfo{author}{Matsuda, S.}, \bibinfo{author}{Ebihara, M.},
  \bibinfo{author}{Noguchi, T.}, \bibinfo{author}{Tsuchiyama, A.},
  \bibinfo{author}{Yurimoto, H.}, \bibinfo{author}{Zolensky, M.E.},
  \bibinfo{author}{Uesugi, M.}, \bibinfo{author}{Shirai, K.},
  \bibinfo{author}{Abe, M.}, \bibinfo{author}{Yada, T.},
  \bibinfo{author}{Ishibashi, Y.}, \bibinfo{author}{Fujimura, A.},
  \bibinfo{author}{Mukai, T.}, \bibinfo{author}{Ueno, M.},
  \bibinfo{author}{Okada, T.}, \bibinfo{author}{Yoshikawa, M.~Kawaguchi, J.},
  \bibinfo{year}{2011}.
\newblock \bibinfo{title}{Irradiation history of itokawa regolith material
  deduced from noble gases in the hayabusa samples}.
\newblock \bibinfo{journal}{Science} \bibinfo{volume}{333},
  \bibinfo{pages}{1128--1131}.
\bibitem[{Pak and Behringer(1993)}]{Pak1993}
\bibinfo{author}{Pak, H.K.}, \bibinfo{author}{Behringer, R.P.},
  \bibinfo{year}{1993}.
\newblock \bibinfo{title}{Surface waves in vertically vibrated granular
  materials}.
\newblock \bibinfo{journal}{Phys. Rev. Lett.} \bibinfo{volume}{71},
  \bibinfo{pages}{1832--1835}.
\bibitem[{Pak et~al.(1995)Pak, Van~Doorn and Behringer}]{Pak1995}
\bibinfo{author}{Pak, H.K.}, \bibinfo{author}{Van~Doorn, E.},
  \bibinfo{author}{Behringer, R.P.}, \bibinfo{year}{1995}.
\newblock \bibinfo{title}{Effects of ambient gases on granular materials under
  vertical vibration}.
\newblock \bibinfo{journal}{Phys. Rev. Lett.} \bibinfo{volume}{74},
  \bibinfo{pages}{4643--4646}.
\bibitem[{Pastor et~al.(2007)Pastor, Maza, Zuriguel, Garcimart\'in and
  Boudet}]{Pastor2007}
\bibinfo{author}{Pastor, J.}, \bibinfo{author}{Maza, D.},
  \bibinfo{author}{Zuriguel, I.}, \bibinfo{author}{Garcimart\'in, A.},
  \bibinfo{author}{Boudet, J.F.}, \bibinfo{year}{2007}.
\newblock \bibinfo{title}{Time resolved particle dynamics in granular
  convection}.
\newblock \bibinfo{journal}{Physica D} \bibinfo{volume}{232},
  \bibinfo{pages}{128 -- 135}.
\bibitem[{Richardson et~al.(2004)Richardson, Melosh and
  Greenberg}]{Richardson2004}
\bibinfo{author}{Richardson, J.E.}, \bibinfo{author}{Melosh, H.J.},
  \bibinfo{author}{Greenberg, R.}, \bibinfo{year}{2004}.
\newblock \bibinfo{title}{Impact-induced seismic activity on asteroid 433 eros:
  A surface modification process}.
\newblock \bibinfo{journal}{Science} \bibinfo{volume}{306},
  \bibinfo{pages}{1526--1529}.
\bibitem[{Richardson~Jr et~al.(2005)Richardson~Jr, Melosh, Greenberg and
  O'Brien}]{Richardson2005}
\bibinfo{author}{Richardson~Jr, J.E.}, \bibinfo{author}{Melosh, H.J.},
  \bibinfo{author}{Greenberg, R.J.}, \bibinfo{author}{O'Brien, D.P.},
  \bibinfo{year}{2005}.
\newblock \bibinfo{title}{The global effects of impact-induced seismic activity
  on fractured asteroid surface morphology}.
\newblock \bibinfo{journal}{Icarus} \bibinfo{volume}{179}, \bibinfo{pages}{325
  -- 349}.
\bibitem[{Saito et~al.(2006)Saito, Miyamoto, Nakamura, Ishiguro, Michikami,
  Nakamura, Demura, Sasaki, Hirata, Honda, Yamamoto, Yokota, Fuse, Yoshida,
  Tholen, Gaskell, Hashimoto, Kubota, Higuchi, Nakamura, Smith, Hiraoka, Honda,
  Kobayashi, Furuya, Matsumoto, Nemoto, Yukishita, Kitazato, Dermawan, Sogame,
  Terazono, Shinohara and Akiyama}]{Saito2006}
\bibinfo{author}{Saito, J.}, \bibinfo{author}{Miyamoto, H.},
  \bibinfo{author}{Nakamura, R.}, \bibinfo{author}{Ishiguro, M.},
  \bibinfo{author}{Michikami, T.}, \bibinfo{author}{Nakamura, A.M.},
  \bibinfo{author}{Demura, H.}, \bibinfo{author}{Sasaki, S.},
  \bibinfo{author}{Hirata, N.}, \bibinfo{author}{Honda, C.},
  \bibinfo{author}{Yamamoto, A.}, \bibinfo{author}{Yokota, Y.},
  \bibinfo{author}{Fuse, T.}, \bibinfo{author}{Yoshida, F.},
  \bibinfo{author}{Tholen, D.J.}, \bibinfo{author}{Gaskell, R.W.},
  \bibinfo{author}{Hashimoto, T.}, \bibinfo{author}{Kubota, T.},
  \bibinfo{author}{Higuchi, Y.}, \bibinfo{author}{Nakamura, T.},
  \bibinfo{author}{Smith, P.}, \bibinfo{author}{Hiraoka, K.},
  \bibinfo{author}{Honda, T.}, \bibinfo{author}{Kobayashi, S.},
  \bibinfo{author}{Furuya, M.}, \bibinfo{author}{Matsumoto, N.},
  \bibinfo{author}{Nemoto, E.}, \bibinfo{author}{Yukishita, A.},
  \bibinfo{author}{Kitazato, K.}, \bibinfo{author}{Dermawan, B.},
  \bibinfo{author}{Sogame, A.}, \bibinfo{author}{Terazono, J.},
  \bibinfo{author}{Shinohara, C.}, \bibinfo{author}{Akiyama, H.},
  \bibinfo{year}{2006}.
\newblock \bibinfo{title}{Detailed images of asteroid 25143 itokawa from
  hayabusa}.
\newblock \bibinfo{journal}{Science} \bibinfo{volume}{312},
  \bibinfo{pages}{1341--1344}.
\bibitem[{Scheeres et~al.(2010)Scheeres, Hartzell, S\'anchez and
  Swift}]{Scheeres2010}
\bibinfo{author}{Scheeres, D.}, \bibinfo{author}{Hartzell, C.I.},
  \bibinfo{author}{S\'anchez, P.}, \bibinfo{author}{Swift, M.},
  \bibinfo{year}{2010}.
\newblock \bibinfo{title}{Scaling forces to asteroid surface: The role of
  cohesion}.
\newblock \bibinfo{journal}{Icarus} \bibinfo{volume}{210},
  \bibinfo{pages}{968--984}.
\bibitem[{Taguchi(1992)}]{Taguchi1992}
\bibinfo{author}{Taguchi, Y.h.}, \bibinfo{year}{1992}.
\newblock \bibinfo{title}{New origin of a convective motion: Elastically
  induced convection in granular materials}.
\newblock \bibinfo{journal}{Phys. Rev. Lett.} \bibinfo{volume}{69},
  \bibinfo{pages}{1367--1370}.
\bibitem[{Tancredi et~al.(2012)Tancredi, Maciel, Heredia, Richeri and
  Nesmachnow}]{Tancredi2012}
\bibinfo{author}{Tancredi, G.}, \bibinfo{author}{Maciel, A.},
  \bibinfo{author}{Heredia, L.}, \bibinfo{author}{Richeri, P.},
  \bibinfo{author}{Nesmachnow, S.}, \bibinfo{year}{2012}.
\newblock \bibinfo{title}{Granular physics in low-gravity environments using
  discrete element method}.
\newblock \bibinfo{journal}{Monthly Notices of the Royal Astronomical Society}
  \bibinfo{volume}{420}, \bibinfo{pages}{3368--3380}.
\bibitem[{Tsuchiyama et~al.(2011)Tsuchiyama, Uesugi, Matsushima, Michikami,
  Kadono, Nakamura, Uesugi, Nakano, Sandford, Noguchi, Matsumoto, Matsuno,
  Nagano, Imai, Takeuchi, Suzuki, Ogami, Katagiri, Ebihara, Ireland, Kitajima,
  Nagao, Naraoka, Noguchi, Okazaki, Yurimoto, Zolensky, Mukai, Abe, Yada,
  Fujimura, Yoshikawa and Kawaguchi}]{Tsuchiyama2011}
\bibinfo{author}{Tsuchiyama, A.}, \bibinfo{author}{Uesugi, M.},
  \bibinfo{author}{Matsushima, T.}, \bibinfo{author}{Michikami, T.},
  \bibinfo{author}{Kadono, T.}, \bibinfo{author}{Nakamura, T.},
  \bibinfo{author}{Uesugi, K.}, \bibinfo{author}{Nakano, T.},
  \bibinfo{author}{Sandford, S.A.}, \bibinfo{author}{Noguchi, R.},
  \bibinfo{author}{Matsumoto, T.}, \bibinfo{author}{Matsuno, J.},
  \bibinfo{author}{Nagano, T.}, \bibinfo{author}{Imai, Y.},
  \bibinfo{author}{Takeuchi, A.}, \bibinfo{author}{Suzuki, Y.},
  \bibinfo{author}{Ogami, T.}, \bibinfo{author}{Katagiri, J.},
  \bibinfo{author}{Ebihara, M.}, \bibinfo{author}{Ireland, T.R.},
  \bibinfo{author}{Kitajima, F.}, \bibinfo{author}{Nagao, K.},
  \bibinfo{author}{Naraoka, H.}, \bibinfo{author}{Noguchi, T.},
  \bibinfo{author}{Okazaki, R.}, \bibinfo{author}{Yurimoto, H.},
  \bibinfo{author}{Zolensky, M.E.}, \bibinfo{author}{Mukai, T.},
  \bibinfo{author}{Abe, M.}, \bibinfo{author}{Yada, T.},
  \bibinfo{author}{Fujimura, A.}, \bibinfo{author}{Yoshikawa, M.},
  \bibinfo{author}{Kawaguchi, J.}, \bibinfo{year}{2011}.
\newblock \bibinfo{title}{Three-dimensional structure of hayabusa samples:
  Origin and evolution of itokawa regolith}.
\newblock \bibinfo{journal}{Science} \bibinfo{volume}{333},
  \bibinfo{pages}{1125--1128}.
\bibitem[{Zeilstra et~al.(2008)Zeilstra, Collignon, van~der Hoef, Deen and
  Kuipers}]{Zeilstra2008}
\bibinfo{author}{Zeilstra, C.}, \bibinfo{author}{Collignon, J.},
  \bibinfo{author}{van~der Hoef, M.}, \bibinfo{author}{Deen, N.},
  \bibinfo{author}{Kuipers, J.}, \bibinfo{year}{2008}.
\newblock \bibinfo{title}{Experimental and numerical study of wall-induced
  granular convection}.
\newblock \bibinfo{journal}{Powder Technology} \bibinfo{volume}{184},
  \bibinfo{pages}{166 -- 176}.

\end{thebibliography}
\end{document}